\documentclass[useAMS,usegraphicx]{mn2e}

\title[Blurred X-ray reflection in Ark~120]{A reflection origin for the soft and hard X-ray excess of Ark~120}
\author[E. Nardini et al.]{E.~Nardini,$^{1,2,3}$\thanks{E-mail: emanuele@ast.cam.ac.uk} A.C. Fabian,$^3$ R.C. Reis$^3$ and D.J. Walton$^3$\\
$^1$ Dipartimento di Fisica e Astronomia - Sezione di Astronomia, Universit\`a di Firenze, L.go E. Fermi 2, 50125 Firenze, Italy\\
$^2$ INAF - Osservatorio Astrofisico di Arcetri, L.go E. Fermi 5, 50125 Firenze, Italy\\
$^3$ Institute of Astronomy, Madingley Road, Cambridge CB3 0HA}

\begin{document}

\date{Released Xxxx Xxxxx XX}

\pagerange{\pageref{firstpage}--\pageref{lastpage}} \pubyear{2010}

\maketitle

\label{firstpage}

\begin{abstract}
Over the last few years several models have been proposed to interpret the widespread soft excess observed in 
the X-ray spectra of type~1 active galactic nuclei (AGN). In particular, reflection from the photoionized accretion 
disc blurred by relativistic effects has proven to be successful in reproducing both the spectral shape and the variability 
pattern of many sources. As a further test to this scenario we present the analysis of a recent $\sim$100~ks long 
\textit{Suzaku} observation of Arakelian~120, a prototypical `bare' Seyfert~1 galaxy in which no complex absorption 
system is expected to mimic a soft excess or mask the intrinsic properties of this key component. We show that a reflection 
model allowing for both warm/blurred and cold/distant reprocessing provides a self-consistent and convincing interpretation 
of the broadband X-ray emission of Ark~120, also characterized by a structured iron feature and a high-energy hump. 
Although warm absorbers, winds/outflows and multiple Comptonizing regions may play significant roles in sources with 
more spectral complexity, this case study adds evidence to the presence of blurred disc reflection as a basic component 
of the X-ray spectra of type~1 AGN. 
\end{abstract}


%%%%%%%%%%%%%%%%%%%%%%%%%%%%%%%%%%%%%%%%%%%%%%%%%%%%%%%%%%%%%%%%%
\section{Introduction}

Over the 2--10~keV energy range, the X-ray spectra of type~1 active galactic nuclei (AGN) are usually 
well-reproduced by a power-law component, which is believed to originate from the Comptonization 
of cold seed photons in a hot coronal region above the accretion disc, according to the so-called 
two-phase model (Haardt \& Maraschi 1991, 1993). When extrapolated to the lower energies, the 
simple power-law trend fails to account for the extra emission systematically observed in unabsorbed 
Seyfert galaxies (after Arnaud et al. 1985). Due to its extremely smooth shape, this soft excess 
turns out to be consistent with several physical models. In analogy with the high/soft states of
 galactic black hole binaries (Remillard \& McClintock 2006), dominated by thermal emission from 
the disc, the initial attempts at modelling the AGN soft excess were based on single or multicolour 
blackbody components. Although well-fitting, this thermal continuum requires a temperature far higher 
than that expected for a standard thin disc around a supermassive black hole. Moreover, when tested 
on large AGN samples, the characteristic temperature is found to be remarkably constant over almost 
four orders of magnitude in the black hole mass, in spite of very different Eddington rates (e.g. Crummy 
et al. 2006, and references therein; Miniutti et al. 2009). A thermal soft excess does not even comply with 
the Stefan--Boltzmann law for the largest variations in the overall luminosity (Ponti et al. 2006), and the 
inferred size of the emitting region is usually unreasonably small with respect to the gravitational radius 
($r_\rmn{g} = GM/c^2$). \\
In terms of spectral decomposition, similar results can be obtained by invoking the Comptonization of 
disc photons in a cold and optically thick plasma (Page et al. 2004). This raises questions on the nature 
of the X-ray corona. In the case of two detached Comptonizing regions with different physical properties 
(e.g. Dewangan et al. 2007), most of the limitations of the thermal scenario described above also apply 
to the `soft' corona. On the other hand, a single plasma with hybrid thermal/non-thermal electron distribution 
(Coppi 1999) may give rise to both the soft excess and the hard power-law component. \\
Alternatively, the universal shape of the soft X-ray emission has been linked to atomic physics: Gierli{\'n}ski 
\& Done (2004) suggested that the soft excess is actually a fake continuum component due to a broad  
absorption trough at $\sim$2--5~keV, arising in partially ionized gas along the line of sight subject to high 
velocity smearing. The latest simulations, however, prove that the properties of any realistic accretion disc 
wind are not able to reproduce the observed smoothness of the soft excess (Schurch \& Done 2007; 
Schurch, Done \& Proga 2009). Another viable explanation is that of reflection from the photoionized 
surface layers of the accretion disc itself, where the relativistic motion of the infalling matter provides the 
blurring of the narrow atomic features (Fabian et al. 2002; Crummy et al. 2006). This model occasionally 
implies a strong suppression of the intrinsic power law in order to account for a prominent soft excess, 
as justified in the context of strong gravitational light bending (Miniutti \& Fabian 2004). \\
On sheer statistical grounds all the interpretations outlined so far yield acceptable results (Sobolewska 
\& Done 2007). To pursue further this kind of study, access is needed to energies beyond $\sim$10~keV, 
where the physical models make different predictions. The advent of \textit{Suzaku} (Mitsuda et al. 
2007) and its hard X-ray detector (HXD; Takahashi et al. 2007) has made it possible to put solid spectral 
constraints up to $\sim$70~keV, opening a new era of AGN observations. Concerning the soft excess, 
it should also be noticed that the possible presence of complex and variable absorption can mask the 
intrinsic appearance of this critical component. Indeed, warm absorbers are very common among type~1 
AGN (Crenshaw, Kraemer \& George 2003; Blustin et al. 2005). It is therefore desirable to select a target 
with the cleanest view of the nuclear regions. \\
Arakelian~120 ($z=0.0327$) is a rare case of a `bare' Seyfert galaxy, hence it represents the optimal 
candidate to test whether the soft excess is a signature of blurred reflection from the inner accretion disc. No 
evidence for reddening is found in the infrared (Ward et al. 1987), and ultraviolet observations establish that 
Ark~120 is devoid of intrinsic absorption (Crenshaw et al. 1999). In the soft X-ray band the source has been 
shown to have a steep spectrum by both \textit{EXOSAT} (Turner \& Pounds 1989) and \textit{ROSAT} 
(Brandt et al. 1993), while the \textit{XMM-Newton} Reflection Grating Spectrometer data allow stringent 
upper limits ($\sim$1--2 orders of magnitude lower than those of usual Seyfert~1s) to be placed on the 
ionic column densities of any possible warm absorber (Vaughan et al. 2004). It is worth emphasizing that 
Ark~120 is a broad-line Seyfert 1 (BLS1) galaxy, whereas the soft excess has been long associated with 
narrow-line sources only. NLS1s are rather eccentric objects, known for their very steep X-ray spectrum 
(Boller, Brandt \& Fink 1996), high accretion rate (Grupe 2004, and references therein), large-amplitude X-ray 
variability on short timescales (Gallo et al. 2004), metal overabundance (Shemmer \& Netzer 2002; Fabian 
et al. 2009) and enhanced star formation (Sani et al. 2010). Ark~120 is an outstanding counter example of a 
normal BLS1 with a prominent soft excess, ensuring as such a completely unbiased exploration of this component. \\
This work is organized as follows: Section~2 concerns the observation and data reduction; our results are 
presented and fully discussed in Section~3, while in Section~4 we summarize our conclusion and outline the 
future research. 

%%%%%%%%%%%%%%%%%%%%%%%%%%%%%%%%%%%%%%%%%%%%%%%%%%%%%%%%%%%%%%%%%
\section{Observation and data reduction}

Ark~120 was observed by \textit{Suzaku} on 2007 April 1--3 in the HXD nominal position, with a resulting net exposure of 
$\sim$101~ks for the X-ray imaging spectrometer (XIS; Koyama et al. 2007) and $\sim$89~ks for the HXD/PIN detector. 
Events were collected by the three operational XIS CCDs in both 3x3 and 5x5 editing modes (71 and 30~ks exposures, 
respectively). Following the standard procedure illustrated in the \textit{Suzaku} Data Reduction 
Guide,\footnote{http://heasarc.gsfc.nasa.gov/docs/suzaku/analysis/abc/} we used the \textsc{heasoft ftools} 6.8 package to 
obtain a new set of clean event files for each detector, editing mode and telemetry by reprocessing the unfiltered events with the 
latest calibrations. The source spectra and light curves were extracted from circular regions with radius of $\sim$3.5 arcmin 
(200 pixels) centred on the target, while the background was evaluated on the same chip from adjacent regions devoid of 
significant contamination. Finally, the source and background spectra from the two front-illuminated detectors (XIS0 and XIS3), 
as well as the response files generated through the `xisresp' script, were merged and rebinned by a factor of 4. \\
Concerning the HXD/PIN data reduction, we again reprocessed the unfiltered event files using the standard tools and 
got the output spectrum by running the `hxdpinxbpi' script, which takes into account the contribution of both the non X-ray 
and the cosmic X-ray background and applies the dead time correction. This returned a source count rate of 
$(12.1 \pm 0.3) \times 10^{-2}$~s$^{-1}$, corresponding to 18.7 per cent of the PIN total counts. \\
The spectral analysis has been performed using the \textsc{xspec} v12.6 fitting package, and involves only the 0.5--12~keV 
energy range of the two front-illuminated XIS detectors; indeed, the 1.7--2.0~keV interval appears to be strongly affected by 
systematic calibration uncertainties around the instrumental silicon K-edge, hence it has been excluded. The XIS1 back-illuminated 
spectrum has been employed throughout as an independent check. The source is confidently detected at high energy up 
to $\sim$40--50~keV; we have therefore considered the conservative 12--40~keV range of the HXD/PIN spectrum. In order 
to allow the use of $\chi^2$ minimization during the spectral fitting, the XIS data were grouped so that each energy channel 
contains no less than 20 counts, while a minimum of 50 counts per bin was adopted for the HXD/PIN spectrum. 
The uncertainties reported in this work correspond to the 90 per cent confidence intervals ($\Delta \chi^2 = 2.71$) for the 
single parameter of interest. Fluxes and count rates are given at the 1$\sigma$ level.

%%%%%%%%%%%%%%%%%%%%%%%%%%%%%%%%%%%%%%%%%%%%%%%%%%%%%%%%%%%%%%%%%
\section{Data analysis and discussion}

\begin{figure}
\includegraphics[width=8.5cm]{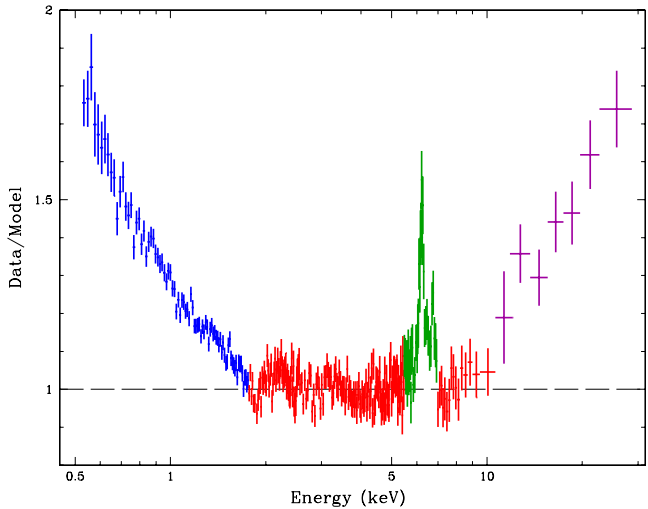}
\caption{The \textit{Suzaku} XIS0+XIS3 and PIN spectra of Ark~120 plotted as a data/model ratio after fitting the 2.5--5.5~keV range only 
with a power law. This sketch, though model-dependent, is straightforward in bringing out both the prominent iron feature at 
$\sim$6~keV and the overall spectral curvature, the latter suggesting the presence of a smooth soft excess below 2~keV 
and of a broad reflection hump beyond 10~keV. (The spectral data shown throughout this work have been rebinned for 
plotting purposes only).}
\label{rp}
\end{figure}

In Fig.~\ref{rp} we show the data/model ratio plot obtained by fitting the \textit{Suzaku} spectrum of Ark~120 with a simple 
power law over the 2.5--5.5~keV range. The output is obviously model-dependent, especially with respect to the magnitude 
of the soft and hard excess; none the less, both the sizable spectral curvature and the prominent iron emission are clearly 
brought out. Our analysis is first aimed at a detailed inspection of the latter feature, and is subsequently extended to the 
whole 0.5--40~keV energy range. Also, we briefly examine the source variability, and review the previous \textit{XMM-Newton} 
observation of Ark~120 in the light of the new evidence from the present work. 

\subsection{Iron K-shell emission}

After excluding the 5--7.5~keV region characterized by the complex iron feature, the 2.5--12~keV spectrum has 
been fitted with a basic model consisting of a power law modified by Galactic absorption only. Despite the 
apparent lack of local obscuration in Ark~120, the foreground column density towards the source is quite large 
due to its low Galactic latitude ($b=-21 \fdg 13$). Two different values for $N_\rmn{H}$ have been determined from 
the $\lambda$21-cm neutral hydrogen emission maps, i.e. $0.98 \times 10^{-21}$~cm$^{-2}$ (Kalberla et al. 2005) 
and $1.26 \times 10^{-21}$~cm$^{-2}$ (Dickey \& Lockman 1990). Throughout this work, Galactic absorption has 
been modelled by means of the \textsc{tbabs} code (Wilms, Allen \& McCray, 2000), assuming the estimates above to be 
the extremes of the allowed range for $N_\rmn{H}$. We will briefly return to this point later on, since for now the selection of 
a precise value of $N_\rmn{H}$ does not affect the study of iron emission around 6~keV. \\
\begin{figure}
\includegraphics[width=8.5cm]{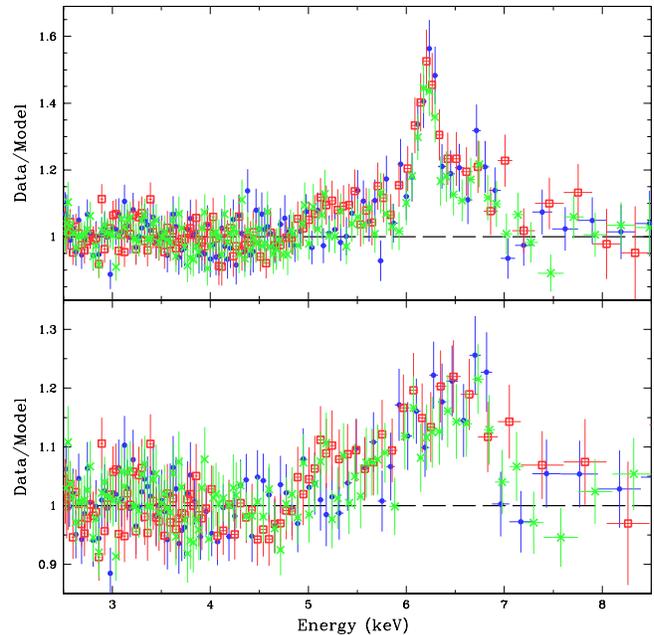}
\caption{Top panel: the structured profile of iron K-shell emission obtained as a data/model ratio after fitting the 
2.5--5 plus 7--12~keV energy range with a single power law. It is evident that two narrow gaussian components 
are not sufficient to reproduce the whole feature. The data from XIS0 (blue dots), XIS1 (red squares) and XIS3 
(green crosses) are shown separately. Bottom panel: same as above, after removing the disc line component 
from the best fitting-model.}
\label{ka}
\end{figure}
The shape of this feature is shown in the top panel of Fig.~\ref{ka}. Different attempts have been made in order to 
disentangle the components of such a structured profile. Here we consider the following four cases: two narrow lines; 
three narrow lines; two narrow lines and a disc line; two broad lines. \\
1) A model including only a couple of narrow gaussian lines of width $\sigma = 10$~eV yields a statistically good fit to 
the data, with $\chi^2_\nu/\rmn{d.o.f.}=0.869/283$. The lines are centred at $\sim$6.42 and 6.96~keV in the source frame, 
and are consistent with the energy expected for K-shell emission from neutral and H-like iron; their equivalent widths (EW) 
are $\sim$90 and 40~eV, respectively. However, this solution is clearly insufficient to reproduce the observed feature. \\
2) By adding another narrow line, the fit undergoes a considerable improvement ($\chi^2_\nu/\rmn{d.o.f.}=0.789/281$). 
The new line has an energy of $\sim$6.67~keV, driving a straightforward identification as He-like iron emission, and an 
EW of $\sim$30~eV. But also this interpretation is not fully adequate. \\
3) In fact, the most successful way to fit the residuals left by the two narrow lines is to allow for a disc line, whose broadening 
is due to relativistic effects. By including in the model a \textsc{laor} profile (Laor 1991) we get an excellent fit of the whole iron 
emission feature ($\chi^2_\nu/\rmn{d.o.f.}=0.686/280$). The shape of this skewed component is shown in the bottom panel of 
Fig.~\ref{ka}, and clearly exhibits the red elongated wing typical of lines arising from the inner disc, even if not as prominent as 
in the well-known case of MCG--6-30-15 (Miniutti et al. 2007). For this reason, it is not possible to constrain simultaneously all 
the line parameters (see Table~\ref{t1}): these include the disc inner radius ($r_\rmn{in}$) and inclination with respect to the line 
of sight and the emissivity index $q$ (under the assumption of a radial emissivity profile $\epsilon (r) \propto r^{-q}$). None the 
less, we note that the best-fitting inner radius of $\sim$13~$r_\rmn{g}$ does not necessarily require the extreme gravity regime 
typical of rapidly rotating black holes. It is important to stress that alternative explanations invoking a spectral upturn due to 
complex absorption effects, such as those proposed for the same MCG--6-30-15, are definitely not viable in the case of Ark~120. \\
4) In the latter stage, the properties of the two narrow lines are exactly the same as found in the original model. Thus, it is worth 
investigating another possibility by retaining the basic template and allowing the width of the two lines to vary. The consequent 
fit refinement is not as significant as when a disc line is involved, being now $\chi^2_\nu/\rmn{d.o.f.}=0.735/282$: the 
difference of $\Delta \chi^2 \simeq -15.2$ obtained with the loss of two degrees of freedom is not likely to be 
simply a chance improvement and supports the relativistic line detection. Moreover, although the double-peaked energy of the 
blended feature is still consistent within the errors with neutral and H-like iron emission (Table~\ref{t1}), the resulting width of 
$\sigma = 113 (\pm 21)$~eV poses the question about the physical location wherein these lines arise. Such a value, in fact, 
corresponds to a full width at half-maximum (FWHM) broadening of $\sim 12 \times 10^3$~km~s$^{-1}$, which is a factor of 
$\sim$2 larger than that observed in the optical permitted lines (FWHM H$\beta$ $\simeq 5800$~km~s$^{-1}$; Wandel, Peterson 
\& Malkan 1999). This could hint at a sort of X-ray broad-line region (BLR) internal to the optical one. \\
Any further discussion on the possible origin of these lines (either broad or narrow) is deferred to the next section, in which 
we address this issue within the context of X-ray reflection models. We are confident that relativistic effects are in place
 and have to be taken into account also when considering the entire spectral range, since all the interpretations 
of the iron emission profile in Ark~120 making no resort to a disc line turn out to be less successful and convincing. 
\begin{table*}
\caption{Best-fitting parameters for different models of the iron K-shell feature: 1) two narrow lines; 2) three narrow lines; 
3) two narrow lines plus a \textsc{laor} disc line; 4) two broad lines. The asterisk-marked parameters have not been 
allowed to vary.}
\label{t1}
\begin{tabular}{l@{\hspace{15pt}}c@{\hspace{25pt}}l@{\hspace{15pt}}c}
\hline
$E_1{}^a$ & $6.42^{+0.02}_{-0.01}$ & EW$_1{}^b$ & $89 \pm 20$ \\
$E_2{}^a$ & $6.96 \pm 0.03$ & EW$_2{}^b$ & $41 \pm 12$ \\
$\sigma{}^b$ & $10^*$ & $\chi^2_\nu$  & 246/283 \\
\hline
$E_1{}^a$ & $6.42^{+0.01}_{-0.02}$ & EW$_1{}^b$ & $86 \pm 20$ \\
$E_2{}^a$ & $6.67^{+0.04}_{-0.05}$ & EW$_2{}^b$ & $26 \pm 17$ \\
$E_3{}^a$ & $6.97 \pm 0.03$ & EW$_3{}^b$ & $41 \pm 22$ \\
$\sigma{}^b$ & $10^*$ & $\chi^2_\nu$  & 222/281 \\
\hline
$E_1{}^a$ & $6.42^{+0.01}_{-0.02}$ & EW$_\rmn{L}{}^b$ & $123^{+77}_{-65}$ \\
$E_2{}^a$ & $6.98^{+0.04}_{-0.03}$ & $q{}^c$ & $3.0^*$ \\
$\sigma{}^b$ & $10^*$ & $r_\rmn{in}{}^d$ & $13^{+19}_{-7}$ \\
EW$_1{}^b$ & $61^{+25}_{-23}$ & $r_\rmn{out}{}^d$ & $400^*$ \\
EW$_2{}^b$ & $43^{+25}_{-24}$ & $\theta{}^e$ & $40^*$ \\
$E_\rmn{L}{}^a$ & $6.46^{+0.06}_{-0.07}$ & $\chi^2_\nu$ & 192/280 \\
\hline
$E_1{}^a$ & $6.42 \pm 0.02 $ & EW$_1{}^b$ & $137 \pm 33$  \\
$E_2{}^a$ & $6.94^{+0.04}_{-0.05}$ & EW$_2{}^b$ & $64 \pm 26$ \\
$\sigma{}^b$ & $113 \pm 21 $ & $\chi^2_\nu$  & 207/282 \\
\hline
\end{tabular}
\flushleft
$^a$ Line peak energy, in keV. \\
$^b$ Line (equivalent) width, in eV. \\
$^c$ Disc emissivity index, $\epsilon (r) \propto r^{-q}$. \\
$^d$ Disc inner/outer radius, in $r_\rmn{g}$ units. \\
$^e$ Disc inclination, in degrees. \\
\end{table*}

\subsection{Blurred reflection model}

\begin{figure}
\includegraphics[width=8.5cm]{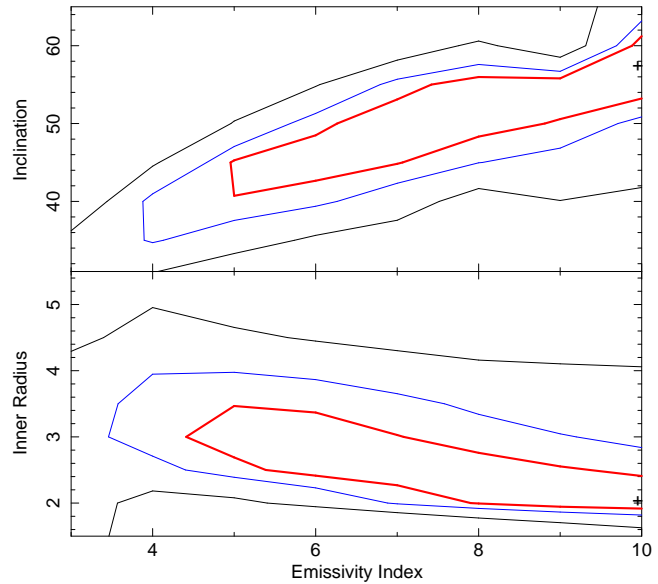}
\caption{Contour plots (68, 90, 99 per cent confidence levels) showing the degeneracy among the critical blurring 
parameters. The emissivity index (horizontal axis) appears to be poorly constrained against the disc inclination 
(vertical axis, upper panel) and inner radius (vertical axis, lower panel). The crosses mark the best-fitting solution 
when all the three parameters are free to vary.}
\label{deg}
\end{figure}

We now extend our analysis to the whole 0.5--40~keV energy range, in order to understand the origin of the 
spectral curvature.\footnote{The cross calibration between the XIS and PIN spectra has been fixed to the 
recommended value of 1.17.} Excess emission beyond $\sim$20--30~keV is usually interpreted as due to the reprocessing 
of the primary X-ray radiation: the combination of photoelectric absorption and Compton scattering in the illuminated 
material gives rise to a broad reflection hump (e.g. George \& Fabian 1991, and references therein). Besides this 
additional continuum component and iron fluorescence, below $\sim$2~keV the reflected spectrum is expected to 
be dominated by a wealth of emission lines from oxygen and other abundant elements, like C, N, Ne, Mg, Si, S 
(Ross \& Fabian 1993). Since it is fairly conceivable that in many cases the accretion flow itself acts as the most 
efficient \textit{mirror}, depending on the ionization stage of the disc outer layers and on the relativistic motions of 
the inner regions, the stack of individual features can be blurred into the smooth shape of the soft excess. \\
As stated above, the slight spectral curvature observed at $\sim$6~keV in Ark~120 cannot be explained as the 
product of (multiple) covering effects, suggesting instead the presence of a broad skewed profile within the iron 
emission feature and indicating a strong gravity regime. It is therefore reasonable to include in our general model two 
reflection components, for which we have used the self-consistent \textsc{reflionx} table models of Ross \& Fabian (2005): 
the first one is expected to arise from almost neutral material at great distance from the X-ray emitting region, likely situated 
on the dusty torus scale. The second one can be ascribed to the partially ionized surface of the disc, and has been 
convolved with the \textsc{kdblur} kernel in \textsc{xspec} to account for relativistic effects. The model obviously comprises 
the primary power law, absorbed by the Galactic column density, and also the two unresolved gaussian lines identified 
above, corresponding to neutral and H-like iron emission. Galactic absorption is found to be consistent with 
$N_\rmn{H}=0.98 \times 10^{-21}$~cm$^{-2}$, and has been frozen accordingly. \\
\begin{figure}
\includegraphics[width=8.5cm]{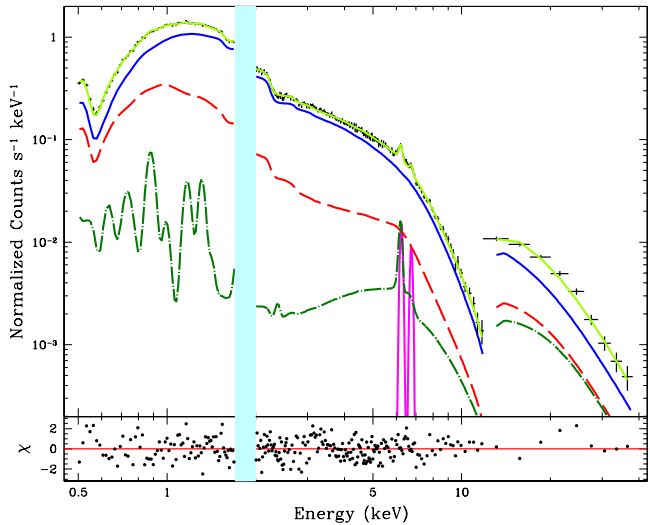}
\caption{Top panel: \textit{Suzaku} XIS and HXD/PIN spectrum of Ark~120 and best-fitting reflection model (light green 
solid line). The contribution of all the different components is disentangled: primary power law (blue), blurred (red, dashed) 
and distant (deep green, dot-dashed) reflection, narrow iron lines (magenta). The shaded 1.7--2~keV region has been 
excluded from the fit. Bottom panel: residuals in units of $\sigma$ (the error bars have size one).}
\label{bf}
\end{figure}
The reflection templates include among their free parameters iron abundance (which has been assumed to be 
single-valued across the source), ionization state (described by the quantity $\xi = 4 \pi F/n_\rmn{H}$, where 
$n_\rmn{H}$ is the hydrogen number density of the gas exposed to an X-ray flux $F$) and photon index of the 
illuminating power law (which has been taken to be one and the same with the direct power law). The key 
blurring parameters are those already introduced above for the \textsc{laor} shape, i.e. inner radius $r_\rmn{in}$, 
emissivity index $q$ and disc inclination. The disc outer radius is usually hard to constrain and has been fixed to 
400~$r_\rmn{g}$. The normalizations of the power-law and reflection components are not tied to each other: in fact, 
in the presence of light bending effects (e.g. Miniutti \& Fabian 2004) which may focus the illuminating radiation on to 
the disc violating isotropy, the traditional reflection fraction has no clear link with the geometrical covering factor. As a 
consequence, in this work we simply weight the contribution of the reflection components by measuring the ratio 
between the reflected and total observed flux. \\
The blurred reflection model illustrated so far is able to reproduce all the spectral complexity of Ark~120 leaving no clear 
structure in the residuals (see Fig.~\ref{bf}). However, it is necessary to discuss in more detail the best-fitting parameters 
that come to light (Table~\ref{t2}). In particular, the disc emissivity profile is suggested to be very steep ($q>6.2$, 
consistent with the largest value allowed in the model, which is 10). This would imply that energy dissipation is 
extremely concentrated in the disc inner regions, virtually all the emission arising within $\sim$2--3~$r_\rmn{g}$. 
Also, an inclination of 57\degr ~seems to be too large for a source with such a clean line of sight, although 
a misalignment of the accretion disc with the obscuring material is possible (see Lawrence \& Elvis 2010).\footnote{We note 
that the host of this Seyfert nucleus is a low-inclination ($i \sim 26\degr$) spiral galaxy (Nordgren et al. 1995).} 
Indeed, it turns out that at this level of signal-to-noise the critical blurring parameters can be highly degenerate with 
each other, as shown in Fig.~\ref{deg}. Specifically, the emissivity index appears to be poorly constrained with 
respect to the inner radius and the disc inclination. We have therefore performed again all the fitting procedure, 
after freezing in turn each one of these three key parameters to a more convincing value, that is $q=5$, 
$r_\rmn{in} = 3 \ r_\rmn{g}$ and $\theta = 40\degr$ (all obtained from the intermediate analytical steps). In all of these 
attempts we are able to recover a fit that is statistically equivalent to the one of reference: according to an $F$-test, the 
ancillary fits match the general case at a confidence level of 26, 15 and 31 per cent, respectively.\footnote{The 
caveats against the use of the $F$-test for X-ray spectral analysis (e.g. Protassov et al. 2002) mainly concern the 
detection of marginal lines. This is not the case here, and the $F$-test is only intended to assess the degree of degeneracy 
among the key parameters.} In the light of these considerations, we can conclude that the blurred reflection model 
provides a valuable interpretation of the broadband X-ray emission of Ark~120 without requiring any extreme parameter. \\
Owing to the degeneracy discussed above, it is not possible to put solid constraints on the black hole spin, as 
recently done e.g. in the case of Fairall 9 (Schmoll et al. 2009). In this effort we have switched to a different relativistic kernel 
(\textsc{kerrconv}; Brenneman \& Reynolds 2006), which allows the dimensionless spin parameter $a=cJ/GM^2$ to vary, 
assuming that the disc extends down to the innermost stable circular orbit ($J$ and $M$ are the black hole angular momentum 
and mass). The former $q$--$r_\rmn{in}$ degeneracy is shifted into a $q$--$a$ degeneracy; yet, by freezing again $q=5$  
we find that $0.24 < a < 0.93$ at a 99 per cent confidence level, with a best estimate of $a \simeq 0.74$. It should be 
noted that the blurring parameters required to model properly the soft excess do not show an obvious agreement with 
those fitting separately the broad iron line through the same \textsc{laor} profile, as pointed out in the previous section. 
In the framework of a pure disc reflection model, a substantial discrepancy in the blurring/ionization parameters of the 
soft excess and the companion broad iron K line is not expected. Anyway, we remark that the critical quantities 
listed in Table~\ref{t2} should not be taken at their face value, due to the large degeneracy. Indeed, even after freezing 
$r_\rmn{in} = 10 \ r_\rmn{g}$ we still obtain a fully acceptable fit\footnote{Interestingly, in this case the disc 
inclination is $\sim$25\degr, in excellent agreement with that of the host galaxy. The emissivity index is steep ($\sim$7.9) 
but again is poorly constrained.} with $\chi^2_\nu \simeq 0.89$. In addition, it has already been suggested that the line 
detection can be difficult with present data quality unless some favourable conditions are met, e.g. a high iron abundance 
and/or a reflection-dominated spectrum (as for 1H0707--495; Zoghbi et al. 2010). By inspecting carefully the residuals in 
the $\sim$5--7.5~keV energy range, tentative evidence of some unfitted structure is found. We have therefore frozen the 
best-fitting model altogether and tried to add a gaussian profile, realizing that the residuals are significantly smoothed after 
the marginal detection of a narrow line centred at 6.69($\pm 0.06$)~keV, whose EW is $<29$~eV. If included in the model 
from the beginning, this line cannot be distinguished with such precision.\footnote{Moreover, even if a third narrow line is 
added to the best fit of the iron feature alone obtained before, the fit does not improve with any statistical significance.} 
Using Fig.~\ref{ka} as a reference, it seems plausible that the iron K-shell feature in Ark~120 consists of three narrow 
components (due to neutral, He-like and H-like iron) plus a faint relativistic disc line, whose parameters cannot actually 
be constrained against the blurring of the soft excess. \\
It is significant that the narrow lines are not accounted for by the self-consistent reflection model we have adopted. 
This notwithstanding, an origin different than fluorescence in material exposed to intense X-ray illumination is difficult 
to consider. We stress that only the putative H-like line keeps unchanged the properties (i.e. a central energy of 
$\sim$6.97~keV and an EW of $\sim$30~eV) already found in the analysis restricted to the iron region. Conversely, 
the line at lower energy is no longer fully consistent with neutral iron emission (it peaks at $\sim$6.46~keV), and its 
EW is significantly reduced ($\sim$38~eV). This is because genuine neutral iron emission is now embodied in the cold reflection 
component (see Fig.~\ref{bf}). These two lines then point to intermediate/high ionization states of iron, and may arise in a much 
more internal environment. In order to probe this possible location, we have again left the width of these lines free, achieving 
a marginal improvement ($\Delta \chi^2 \simeq -3.7$ with the loss of one degree of freedom, so that the probability 
of chance improvement is non-negligible). All the previous best-fitting parameters are remarkably confirmed, and the 
resulting width of $\sigma \simeq 80 (\pm 50)$~eV corresponds to a FWHM velocity of $8.7 (\pm 4.3) \times 10^3$~km~s$^{-1}$, 
which is consistent with an origin in (or slightly internal to) the optical BLR. The emission lines seem therefore to be connected 
to reflection neither at the pc nor at the $r_\rmn{g}$ scale. A possible explanation hints at some blobs of material orbiting 
in the outskirts of the disc, forming a sort of X-ray BLR as suggested by Blustin \& Fabian (2009), or where the height-radius 
relation of the disc changes. \\
Finally, we briefly comment on the relative amplitudes of the reflection components, which have been computed in terms 
of fractional contributions to the observed flux in both the XIS and the HXD energy range. Shortward of $\sim$10~keV, the cold 
reflector accounts for $\sim$4 per cent of the received emission, its significance growing at higher energies where 
the Compton hump comprises $\sim$18 per cent of the total flux (or equivalently $\sim$30 per cent with 
respect to the power-law component only). This strength can be compared to that expected from a slab subtending 
a solid angle of 2$\pi$ at the illuminating source by performing a straightforward test with a simple \textsc{powerlaw+pexrav} 
model (Magdziarz \& Zdziarski 1995) in \textsc{xspec}. Recalling that \textsc{pexrav} includes only the reflected 
continuum, hence a sound comparison with \textsc{reflionx} is possible only beyond $\approx$2~keV, we find a 
remarkable agreement between our measure and the `dummy' estimate. It is worth noting that even in the presence of 
strong light bending effects in the inner regions of the disc, at large distance from the X-ray source isotropy should be 
recovered: it is therefore desirable for the cold reflector identifiable with the obscuring torus to see the same illuminating 
radiation as the observer at infinity. Such a prescription is entirely complied with in this case. The fractional contribution 
of the warm reflection component from the disc is instead rather flat, increasing with energy from $\sim$19 at 0.5--12~keV 
to only $\sim$22 per cent at 12--40~keV. Again, these values do not pose severe problems in terms of the required reflection 
efficiency below $\sim$10~keV, due to the wealth of soft X-ray emission lines that are not implemented in the standard 
\textsc{pexrav}/\textsc{pexriv} models; besides this, it is reasonable to allow for moderate light bending involving the blurred 
component. \\
In conclusion, a model consisting of both a warm reflection component arising from the photoionized disc surface, blurred 
by relativistic effects, and a cold one ascribable to a much more distant reprocessor proves to be successful in 
reproducing the broadband X-ray spectrum of Ark~120. Limited to this peculiar source, this scenario stands as the most 
convincing among those presented so far, also due to the minimal set of geometrical and physical assumptions involved. 

\begin{table*}
\caption{Best-fitting parameters for the blurred reflection model of the whole 0.5--40~keV \textit{Suzaku} spectrum.}
\label{t2}
\begin{tabular}{l@{\hspace{15pt}}c@{\hspace{25pt}}l@{\hspace{15pt}}c}
\hline
$\Gamma{}^a$ & $2.030^{+0.011}_{-0.004}$ & $\xi_\rmn{b}{}^d$ & $278^{+35}_{-27}$ \\
$E_1$ & $6.46^{+0.02}_{-0.03}$ & $q$ & $> 6.2$ \\
$E_2$ & $6.97 \pm 0.04$ & $r_\rmn{in}$ & $2.04^{+1.58}_{-0.27}$ \\
$\sigma$ & $10^*$ & $\theta$ & $57^{+5}_{-12}$ \\
EW$_1$ & $38^{+32}_{-28}$ & $R_\rmn{d}{}^e$ & $0.034^{+0.003}_{-0.004}$ \\
EW$_2$ & $32^{+11}_{-12}$ & $R_\rmn{b}{}^e$ & $0.178^{+0.030}_{-0.015}$ \\
Fe${}^b$ & $0.73^{+0.14}_{-0.06}$ & $F_\rmn{obs}{}^f$ & $5.29^{+0.28}_{-0.08}$ \\
$\xi_\rmn{d}{}^c$ & $< 10.4$ & $\chi^2_\nu$ & 410.4/468 \\
\hline
Fe & $0.77^{+0.09}_{-0.10}$ & $r_\rmn{in}$ & $3.03^{+0.76}_{-0.65}$ \\
$\xi_\rmn{b}$ & $280^{+47}_{-29}$ & $\theta$ & $42^{+5}_{-4}$ \\
$q$ & $5.0^*$ & $\chi^2_\nu$ & 411.5/469 \\
\hline
Fe & $0.77 \pm 0.13$ & $r_\rmn{in}$ & $3.00^*$ \\
$\xi_\rmn{b}$ & $278^{+45}_{-29}$ & $\theta$ & $44^{+3}_{-6}$ \\
$q$ & $5.7^{+2.6}_{-1.9}$ & $\chi^2_\nu$ & 411.3/469 \\
\hline
Fe & $0.77^{+0.13}_{-0.09}$ & $r_\rmn{in}$ & $3.30^{+0.33}_{-0.61}$ \\
$\xi_\rmn{b}$ & $279^{+46}_{-28}$ & $\theta$ & $40^*$ \\
$q$ & $4.8^{+1.5}_{-1.0}$ & $\chi^2_\nu$ & 412.3/469 \\
\hline
\end{tabular}
\flushleft
$^a$ Photon index. \\
$^b$ Iron abundance, in solar units. \\
$^c$ Ionization parameter of the distant reflector, in erg~cm~s$^{-1}$. \\
$^d$ Ionization parameter of the blurred reflector, in erg~cm~s$^{-1}$. \\
$^e$ Reflection fraction over the 0.5--10~keV range. \\
$^f$ Observed 0.5--10~keV flux, in 10$^{-11}$~erg~cm$^{-2}$~s$^{-1}$. \\
\end{table*}

\subsection{Alternative models}

As we have already mentioned in the introduction, several other models have been adopted in the last decade to fully 
reproduce the high-quality $\sim$0.5--10~keV spectra of AGN provided by \textit{Chandra} and/or \textit{XMM-Newton}, 
specifically concerning the soft excess component. Over the years, most of these interpretations have been either 
discarded as too phenomenological or disputed because of some physical limitations. Nevertheless, for the sake of a 
fitting comparison extending up to $\sim$40~keV against the blurred reflection model, we have also explored some of 
the most common alternatives, on which we give a brief report in this section. All the models listed below differ from 
that based on blurred reflection only in the ingredients apt to describe the soft excess. In particular, they all require a 
cold reflection component with iron abundance $\sim$1.1--1.5 ($\pm 0.4$), accounting for $\sim$4--6 and $\sim$20--30 
per cent of the XIS and HXD/PIN observed flux, respectively, and also for the narrow emission feature at 6.4~keV. 
Other two unresolved gaussian lines are always needed in the 6.47--6.70 and 6.93--7.01~keV range: even if loosely 
consistent with He-like iron emission, the former is apparently trying to compensate somehow for the lack of the 
broadened disc line. In the following, we just provide the key parameters of the soft excess component in some illustrative cases. \\
1) \textit{Blackbody} ($\chi^2_\nu/\rmn{d.o.f.}=0.995/471$): the best-fitting disc temperature is $\sim$0.14~keV, and 
therefore lies in the range characteristic to the thermal scenario. No significant improvement can be obtained by adopting 
a more realistic (e.g. multicolour) blackbody profile. \\
2) \textit{Broken power law} ($\chi^2_\nu/\rmn{d.o.f.}=0.963/471$): as one can immediately see from Fig.~\ref{rp}, a 
double-sloped power law is sufficient to achieve a good fit to the low-energy spectrum. The soft photon index is $\sim$2.33, 
the hard one is $\sim$2.00; the break-point energy is established at 1.73~keV. \\
3) \textit{Smeared absorption} ($\chi^2_\nu/\rmn{d.o.f.}=1.02/470$): we have used the \textsc{swind1} model to mimic 
absorption in a partially ionized relativistic gas, even if the latest simulations of the velocity and density structure of any 
possible accretion disc wind or outflow rule out this scenario as the origin of a smooth soft excess (Schurch et al. 2009). 
The column denisty of the putative absorber is $N_\rmn{H} \simeq 17 \times 10^{22}$~cm$^{-2}$, and the ionization 
parameter is $\log \xi \simeq 3.3$. The gaussian velocity dispersion in this warm gas is the maximum allowed, i.e. $v/c=0.5$. 
Since the power-law index is slightly steeper than usual, being $\sim$2.12, this model presents larger deviations from the 
data at the higher energies. \\
4) \textit{Cold Comptonization} ($\chi^2_\nu/\rmn{d.o.f.}=0.918/470$): many codes with a different degree of complexity 
are available to describe Comptonization (e.g. Poutanen and Svensson 1996). However, it is not possible to put 
firm constraints on the plasma properties and geometrical structure with the present data quality and energy coverage; 
we have therefore adopted the essential \textsc{comptt} code of Titarchuk (1994). The exact temperature of seed 
photons is not very important, provided that its value is reasonably low, so it has been frozen to 50~eV as broadly 
presumable for a source like Ark~120. The cold corona turns out to have an electron temperature of $\sim$0.28~keV 
with an optical depth of $\sim$13, even if the two parameters are expected to be degenerate to some extent. This 
optically thick soft component has a luminosity of $\sim$10$^{43}$~erg~s$^{-1}$, and can be roughly approximated 
by a blackbody. This gives a size of the emitting region of $\sim 4 \times 10^{11}$~cm, whereas 
$r_\rmn{g} \sim 2 \times 10^{13}$~cm (the mass of the black hole in Ark~120 is estimated from  reverberation mapping 
to be $M_\rmn{BH} \simeq 1.5 \times 10^8 M_{\sun}$; Peterson et al. 2004). \\
Summarizing, the Comptonization scenario can be regarded at present as the only physical alternative to the blurred 
reflection model in order to account for the soft excess of Ark~120, but its details are difficult to probe; the argued 
compactness of the soft X-ray emitting region represents a further problem for the nature and the geometry of the 
Comptonizing plasma. 

\subsection{Timing analysis}

\begin{figure}
\includegraphics[width=8.5cm]{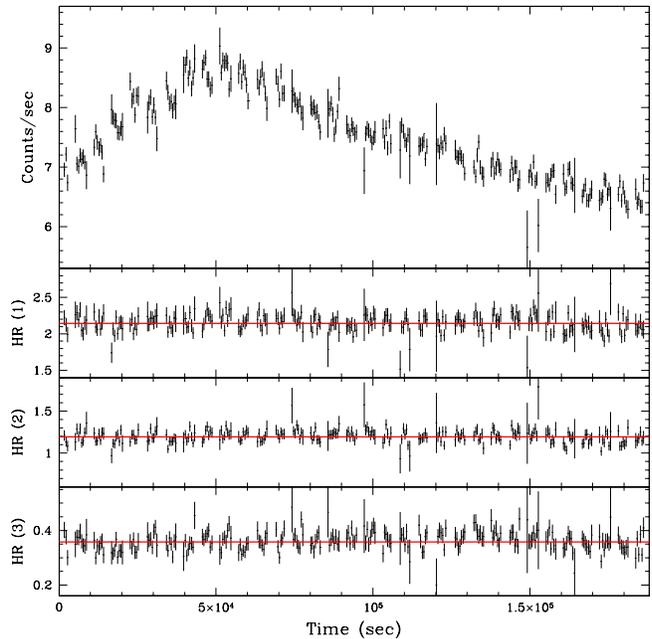}
\caption{Top panel: 0.5--10~keV background-subtracted light curve of Ark~120, summed over the three XIS detectors 
and plotted adopting a time resolution of 500~s. The flux variation is rather gentle, with an overall \textit{rms} variability 
amplitude of $\sim$9 per cent. Lower panels: respectively from above, 1--2, 2--5 and 5--10 over 0.5--1~keV hardness 
ratios, which turn out to be remarkably stable over the entire observation; the red solid lines show the best-fitting constant 
values. (Note: the outlying points with large error bars correspond to the ends of a single \textit{Suzaku} orbit, and are 
just artifacts of the time binning).}
\label{hr}
\end{figure}

In addition to spectral fitting, the other fundamental approach to understand the properties of the central engine in AGN 
is the study of variability. In principle, the latter can also provide independent information to discriminate between the cold 
Comptonization and the blurred reflection scenario for the soft excess, since the two models make quite divergent predictions 
about the behaviour of the source in the time domain. In a very simplistic way, the time lags affecting the different energy 
bands that have been observed in many AGN can be regarded as a natural consequence of a Comptonization process: 
the time delay between the hard and the soft X-ray variations is due to the larger number of scattering events that the 
high-energy photons have to experience before escaping to infinity. However, matters appear to be much more complicated 
than this naive picture (e.g. Ar\'evalo \& Uttley 2006; McHardy et al. 2007). On the other hand, if the soft excess is a reflection 
signature, the soft band should lag behind the hard power-law component as expected in a reverberation context, after 
singling out the timescale corresponding to the light crossing time from the primary source to the reflector and back. Evidence 
in this sense has been recently found in 1H0707--495 (Fabian et al. 2009; Zoghbi et al. 2010). \\
Although \textit{Suzaku} is not an ideal observatory to carry out detailed timing studies of AGN, mainly because of the 
frequent gaps induced by the short orbital period, it is worth checking the variability pattern of Ark~120 for possible hints. 
The total XIS light curve is shown in Fig.~\ref{hr}, and displays a very gentle variation, in part expected due to the large 
black hole mass. Using a time resolution of 500~s, the fractional \textit{rms} variability amplitude $F_\rmn{var}$ (defined 
as in Vaughan et al. 2003) is 9.1($\pm 0.2$) per cent over the 0.5--10~keV range.\footnote{All the values reported 
in this section have little dependence on the adopted time bin. This does not undermine the general conclusions: in this 
source, the light crossing time over a distance of 1~$r_\rmn{g}$ is $\sim$750~s, so that most of the 
variability is expected to take place on larger timescales.} In order to check for a possible energy dependence, we have 
computed the individual variability amplitudes in four energy bands (0.5--1, 1--2, 2--5 and 5--10~keV), which turn out to be 
8.7($\pm 0.4$), 9.7($\pm 0.2$), 9.1($\pm 0.3$) and 8.5($\pm 0.7$) per cent, respectively. This does not provide any straightforward 
indication for different variability patterns, and these are likely to be actually consistent with each other. Moreover, we note that 
the four bands above cannot be clearly linked to different spectral components, since the direct power law is dominant 
over the whole spectrum of Ark~120. Incidentally, the shape of the HXD/PIN light curve closely resembles the trend 
observed in the XIS range, although in this case $F_\rmn{var}=19.1 (\pm 3.4)$, due to the larger noise. A similar indication 
in favour of a coarse coherence with no lags among the energy bands comes from the 1--2, 2--5 and 5--10 over 0.5--1~keV 
hardness ratios (Fig.~\ref{hr}), which are all well-fitted with a constant value ($\chi^{2}/\rmn{d.o.f} \simeq 246/231$, 252/231 and 
267/231, corresponding to null hypothesis probabilities of 24, 16 and 5 per cent, respectively). The lack of differential variability 
was also found by De~Marco et al. (2009), who have analyzed the \textit{XMM-Newton} spectrum focusing on the iron K line region. 
The total time elapsed in this \textit{Suzaku} observation is $\sim$185~ks, corresponding to $\sim$1.2 keplerian orbital periods at 
$10 \ r_\rmn{g}$ (e.g. Bardeen, Press \& Teukolsky 1972). It is therefore unlikely to reveal dramatic changes 
on such timescales, despite the small radii involved. \\
In short, the timing analysis of Ark~120 is inconclusive and does not point to any specific physical model. Yet, by 
confirming that the source variations involve the flux amplitude only, it offers substantial support to our previous spectral 
study, given that the average X-ray emission genuinely represents the single physical state in which the source was 
caught during the \textit{Suzaku} observation. 

\subsection{The 2003 \textit{XMM-Newton} observation}

\begin{table*}
\caption{Best-fitting reflection parameters for the 2003 \textit{XMM-Newton} observation.}
\label{t3}
\begin{tabular}{l@{\hspace{15pt}}c@{\hspace{25pt}}l@{\hspace{15pt}}c}
\hline
$\Gamma$ & $2.116^{+0.011}_{-0.009}$ & $\xi_\rmn{b,1}$ & $299^{+20}_{-18}$ \\
$E_1$ & $6.40^{+0.02}_{-0.01}$ & $\xi_\rmn{b,2}$ & $< 13.9$ \\
$E_2$ & $6.64^{+0.03}_{-0.04}$ & $q$ & $5.2^{+0.4}_{-0.3}$ \\
$E_3$ & $7.00 \pm 0.03$ & $r_\rmn{in}$ & $2.26^{+0.11}_{-0.07}$ \\
$\sigma$ & $10^*$ & $\theta$ & $50^*$ \\
EW$_1$ & $47^{+23}_{-22}$ & $R_\rmn{d}$ & $< 0.008$ \\
EW$_2$ & $26 \pm 7$ & $R_\rmn{b,1}$ & $0.218^{+0.013}_{-0.016}$ \\
EW$_3$ & $25 \pm 15$ & $R_\rmn{b,2}$ & $0.062^{+0.010}_{-0.008}$ \\
Fe & $0.75^*$ & $F_\rmn{obs}$ & $6.80^{+0.44}_{-0.18}$ \\
$\xi_\rmn{d}$ & $1.0^*$ & $\chi^2_\nu$ & 2044/1857 \\
\hline
\end{tabular}
\end{table*}

\begin{figure}
\includegraphics[width=8.5cm]{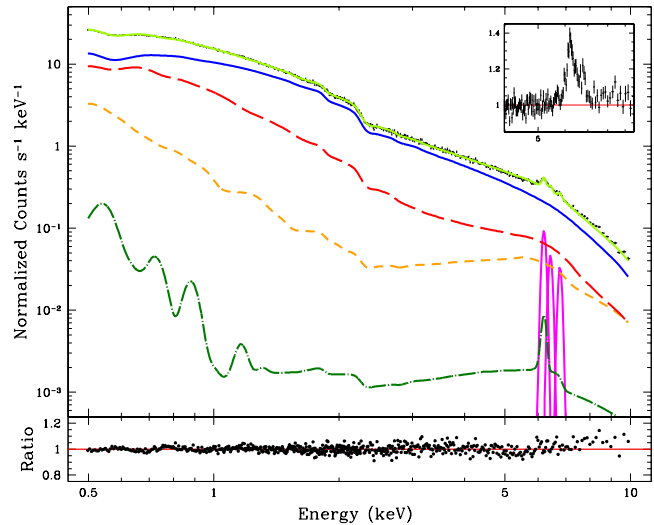}
\caption{Top panel: as Fig.~\ref{bf} but for the 2003 \textit{XMM-Newton} observation, with the same code. The additional 
cold/blurred reflection component is plotted in orange (short-dashed line). The inner box shows the profile of the iron feature, 
obtained as the ratio over a simple power-law trend. Bottom panel: data/model ratio to the best fit, emphasizing the slight 
excess emission longward of the iron feature. Errors bars are omitted for clarity.}
\label{xn}
\end{figure}

Previously to \textit{Suzaku}, Ark~120 was monitored by \textit{XMM-Newton} on 2003 August 24--25. 
A detailed study of this observation is presented in Vaughan et al. (2004). In particular, a wide range of 
models (including reflection components, disc blackbodies, bremsstrahlung and thermal Comptonization) 
were tested in order to give an adequate description of the broadband spectrum; yet, none of these was 
able to account for the properties of the soft excess. It is not our aim to perform the same comprehensive 
analysis over again, reproducing equivalent results. Since then, anyway, the available reflection models 
have been substantially upgraded, after Ross \& Fabian (2005). So it is worth reviewing the very high-quality 
\textit{XMM-Newton} observation in the light of the blurred reflection scenario we have developed in this 
work. Our reduction follows that illustrated in Vaughan et al. (2004), and yields to a full agreement of the 
final data products. The following analysis concerns the EPIC-pn spectrum only, and is confined to the 
0.5--10~keV energy range for consistency with the \textit{Suzaku} XIS operational window. Moreover, 
due to the large count rate in the soft band the ionization and blurring parameters are highly sensitive to 
calibration uncertainties, which can be quite large below $\sim$0.6~keV (see also Papadakis et al. 2010). \\
The spectral shape and flux level in the two observations are remarkably similar, hence we have directly 
applied the best-fitting model obtained for the \textit{Suzaku} case. Iron abundance and disc inclination are 
obviously not time-dependent parameters, and have been frozen to 0.75 (solar units) and 50\degr, respectively. 
Under these assumptions we obtain a reasonable fit to the data, even if not formally acceptable in a statistical 
sense ($\chi^2_\nu \sim 1.16$), as quite usual at the highest levels of data quality. A further improvement can be 
achieved by allowing for a slightly larger degree of complexity in the model. First, a third narrow line is 
included for He-like iron emission as also pointed out by Brenneman \& Reynolds (2009). Second, we 
introduce an additional reflection component, subject to the same relativistic smearing but with an independent 
ionization parameter, in order to allow for possible non-uniformity within the disc. As a result, this extra 
component turns out to require a low ionization, while the cold distant reflection becomes negligible. The value 
of $\chi^2_\nu/\rmn{d.o.f.}$ drops to $\sim$1.10/1857 and the best-fitting key parameters are very well-matched 
to those obtained for the \textit{Suzaku} observation (see Table~\ref{t3}). Conversely, beyond the iron feature there 
remains some excess emission that cannot be fitted by increasing the model complexity. These deviations of 
the model from the data are of the order of $\sim$10 per cent at most (Fig.~\ref{xn}). We note that if the 
disc inclination is left free to vary, a value around $\sim$80\degr ~(see also Nandra et al. 2007) is able to offset 
any $\sim$8--10~keV excess, while the iron abundance is not a critical parameter in this sense. Interestingly, 
the Galactic column density converges towards the upper limit of the admitted range: if instead $N_\rmn{H}$ 
is frozen to the lower value the fit quality marginally worsens, but the free parameters smoothly resettle so 
that the picture above is not modified. We have also checked the behaviour of the model down to 0.3~keV: 
a very clean bump in the data/model ratio emerges at 0.3--0.5~keV. After freezing the best-fitting model, this 
feature is quite well-reproduced by a blackbody of temperature $\sim$40~eV, even if the residuals in the 
0.3--0.7~keV range display some tentative structure. \\
Finally, we have tested for completeness the cold Comptonization model: in this case, the parameters of 
the soft corona are in excellent agreement with those found above for the \textit{Suzaku} spectrum, but the fit is 
less accurate ($\chi^2_\nu \sim 1.19$) and large residuals remain in the 0.5--0.7 and 5.5--7~keV range. Our revision 
of the high-quality \textit{XMM-Newton} observation seems therefore to confirm the reliability of the blurred reflection 
interpretation for soft X-ray emission of Ark~120. 

%%%%%%%%%%%%%%%%%%%%%%%%%%%%%%%%%%%%%%%%%%%%%%%%%%%%%%%%%%%%%%%%%
\section{Conclusions}

We have reported on a long \textit{Suzaku} observation of Ark~120, one of the few known examples of nearly 
absorption-free active nuclei. This `bare' BLS1 galaxy is the optimal candidate to investigate the nature of the 
ubiquitous soft excess observed in the X-ray spectra of unobscured AGN. Its broadband 0.5--40~keV spectrum 
reveals a significant curvature with respect to the main power-law component, as excess emission is found at 
both low ($<2$~keV) and high ($>10$~keV) energies. Also, a complex iron K-shell feature is clearly detected 
at $\sim$5--7~keV, showing a structured profile in which can be identified two (three, tentatively) narrow 
gaussian cores originating roughly at the BLR scale, and a relativistically skewed line arising from the disc 
at $r \ga 10 \ r_\rmn{g}$. The latter, even if not as prominent as in the most impressive cases, cannot be 
interpreted through a blend of unresolved lines or a system of multiple absorbers. Hence relativistic effects have to be 
included in the broadband analysis. \\
We have shown that all the spectral complexity can be successfully explained in terms of a self-consistent reflection 
model, allowing for both a warm/blurred and a cold/distant reflection component. In this picture, the reprocessing of the 
primary X-ray radiation takes place in two distinct regions, that can be plainly identified with the inner accretion disc 
$(r \la 10 \ r_\rmn{g})$ and the far-off obscuring torus. Depending on the exact geometry, however, additional 
physical locations (e.g. a belt of orbiting clouds) may contribute, since fluorescent iron emission suggests the existence 
of a wide range of ionization states. This blurred reflection scenario is further corroborated after reviewing a previous 
high-quality \textit{XMM-Newton} observation of Ark~120 . Therefore, as for this single source, the present interpretation 
stands as the most convincing among those that have been proposed so far, due to the minimal set of geometrical and 
physical assumptions involved. \\
In general it is difficult to rely on comparably clean lines of sight to the nuclear regions, and we expect that other 
processes (likely non-thermal Comptonization, warm absorption, reprocessing in winds/outflows) also play a 
non-negligible role in shaping the X-ray emission of type~1 AGN. As a consequence, our aim in the near future is 
to extend this study to a large sample of low-obscuration sources, possibly observed by both \textit{Suzaku} and 
\textit{XMM-Newton} to combine the high-energy coverage of the former and the large effective area of the latter. 
This will allow us to infer the average contribution of blurred disc reflection to the soft excess and to the X-ray luminosity 
of AGN, and at the same time to test the basic predictions of the light bending model and to explore the nature of 
the illuminating source itself. We note that if blurred reflection is really the dominant component of the soft excess, 
the black hole spin distribution may be biased towards the larger values; indeed, it has already been suggested 
in many previous works (e.g. Volonteri et al. 2005) that the supermassive black holes in AGN should be rapidly 
rotating. Even if the spin cannot be firmly constrained in Ark~120 due to the large degeneracy among the blurring 
parameters (but provisionally $a \simeq 0.7$), our case study is apparently consistent with this scenario. 

%%%%%%%%%%%%%%%%%%%%%%%%%%%%%%%%%%%%%%%%%%%%%%%%%%%%%%%%%%%%%%%%%%%%
\section*{Acknowledgments}

EN acknowledges financial support from the ASI-INAF I/088/06/0 contract. ACF thanks the Royal Society. 
RCR and DJW acknowledge the financial support provided by STFC. The authors are also grateful to the 
anonymous referee for the constructive comments.

%%%%%%%%%%%%%%%%%%%%%%%%%%%%%%%%%%%%%%%%%%%%%%%%%%%%%%%%%%%%%%%%%%%%%%%%%%%

%%%%%%%%%%%%%%%

\label{lastpage}

\end{document}